\documentclass[aps,pre,reprint,superscriptaddress,nofootinbib]{revtex4-1}
\usepackage{graphicx}
\usepackage{float}
\usepackage{csquotes}

\usepackage{mathtools} 

\usepackage[usenames,dvipsnames]{color}
\begin{document}


\title{Non-parametric estimation of a Langevin model driven by correlated noise}


\author{Clemens Willers}
\email[]{clemens.willers@uni-muenster.de}
\thanks{ORCID ID: 0000-0001-5777-0514}
\affiliation{Center for Nonlinear Science (CeNoS), Westf{\"a}lische Wilhelms-Universit\"at M\"unster, Corrensstr.\ 2, 48149 M\"unster, Germany}
\author{Oliver Kamps}
\thanks{ORCID ID: 0000-0003-0986-0878}
\affiliation{Center for Nonlinear Science (CeNoS), Westf{\"a}lische Wilhelms-Universit\"at M\"unster, Corrensstr.\ 2, 48149 M\"unster, Germany}


\date{\today}

\begin{abstract}
Langevin models are frequently used to model various stochastic processes in different fields of natural and social sciences. They are adapted to measured data by estimation techniques such as maximum likelihood estimation, Markov chain Monte Carlo methods, or the non-parametric \textit{direct estimation method} introduced by Friedrich et al. The latter has the distinction of being very effective in the context of large data sets. Due to their $\delta$-correlated noise, standard Langevin models are limited to Markovian dynamics. A non-Markovian Langevin model can be formulated by introducing a hidden component that realizes correlated noise. For the estimation of such a partially observed diffusion a different version of the direct estimation method was introduced by Lehle et al. However, this procedure includes the limitation that the correlation length of the noise component is small compared to that of the measured component. In this work we propose another version of the direct estimation method that does not include this restriction. Via this method it is possible to deal with large data sets of a wider range of examples in an effective way. We discuss the abilities of the proposed procedure using several synthetic examples.
\end{abstract}


\maketitle

\section{Introduction\label{introduction}}

In many different fields of science ranging from turbulence over neuroscience to finance, \textit{Langevin models} are frequently and successfully used to model stochastic processes, see Ref.~\cite{ReviewFriedrichPeinke} for an overview. Such a model is defined by a first-order stochastic differential equation (SDE), the \textit{Langevin equation} \cite{Risken, KloedenPlaten}, which, for a 1-dimensional process $X_t$, reads

\begin{align} \label{LangevinEq}
\dot{X}_t = D^{(1)}(X_t) + \sqrt{D^{(2)}(X_t)}\,\eta_t.
\end{align}

\noindent Hereby, $D^{(1)}$ and $D^{(2)}$ are called \textit{drift function} and \textit{diffusion function}, respectively. The former defines the deterministic part of the dynamics (fixed points, etc.), the latter (which must be non-negative) determines the impact of the \textit{noise} that drives the system. The noise is given by fluctuating Langevin forces $\eta_t$ which realize $\delta$-correlated Gaussian white noise, fulfilling $\langle\eta_t\rangle = 0$ and $\langle\eta_t\eta_{t'}\rangle = \delta(t,t')$.

We do not use a time dependence in the functions $D^{(1)}$ and $D^{(2)}$, i.e., we model stationary processes. The stationarity of a measured time series can be analyzed via a moving-window technique \cite{ReviewFriedrichPeinke}. The same approach can help in the case of a non-stationary process, i.e., different stationary models are estimated for different windows in time.

Numerically, the Langevin equation can be integrated using the \textit{Euler-Maruyama scheme} (we use It\^{o} calculus) \cite{KloedenPlaten}. For a time step $\tau$, the iterative integration is defined by

\begin{align}\label{numericalIntegration}
X_{t+\tau} = X_t + D^{(1)}(X_t)\,\tau + \sqrt{D^{(2)}(X_t)}\,\sqrt{\tau}\, N_t,
\end{align}

\noindent where $N_t$ is an element of a sequence of independent and identically distributed random variables which obey the standard normal distribution: $N_t \sim \mathcal{N}(0,1)$.

The property of the noise to be $\delta$-correlated implies that the process $X_t$ which solves the Langevin equation fulfills the Markov property \cite{Risken}, i.e., there is no memory in its dynamics. The definition of the Markov property reads

\begin{align}\label{Markov}
p(x_{n+1} | x_n, x_{n-1}, ..., x_0) = p(x_{n+1} | x_n)
\end{align}

\noindent where $p(\cdot|\cdot)$ is the conditional probability density function (conditional pdf or cpdf) \cite{Bauer}. Here, we treat the process $X_t$ as discrete in time, because in practice we work with the numerical integration of the Langevin equation or with sampled measurements which both are discrete in time. We use capitals for random variables and lowercases for their realizations.

An important task is the estimation of a suitable Langevin model for a measured time series $(x_0, x_1, x_2, ..., x_N)$. The maximum likelihood estimation \cite{Kleinhans}, Markov chain Monte Carlo methods in a Bayesian framework \cite{Toussaint, BayesianDataAnalysis}, or the \textit{direct estimation method} that was introduced by Friedrich et al. \cite{directEstFriedrich, directEstSiegert} are possible ways to realize such an estimation.

The latter has the distinction of being very simple and computationally cheap. It is very effective, especially in the context of large data sets, and is the estimation method this work is based on. We shall briefly introduce its concept in the following.

The direct estimation method works in a non-parametric way. The values of the measured time series are sorted into bins. Next, averaged values of the drift and diffusion functions in these bins are calculated by means of the first and second moment of the increments of the time series occurring in the corresponding bins.

Precisely, let $(x_0, x_1, x_2, ..., x_N)$ be a measured time series that is sampled with a time step $\tau$. The range of the measured values $D = \lbrack x_{\min}, x_{\max} \rbrack$ is divided into $N_{\text{bins}}$ disjoint parts $B_k$: $D = \cup_{k=1}^{N_{\text{bins}}} B_k$. A part $B_k$ is called bin and centered at $x^k$ (here we use superscripts, subscripts are used for time dependency). For every bin $B_k$, an averaged function value for the drift and diffusion function can be estimated in the following way:

\begin{subequations}
\label{directEst}
\begin{align}
D^{(1)}(x^k) &= \lim_{\tau\to 0} \frac{1}{\tau} \langle x_{i+1} - x_i | x_i\in B_k \rangle\\
D^{(2)}(x^k) &= \lim_{\tau\to 0} \frac{1}{\tau} \Bigl\{ \langle (x_{i+1} - x_i)^2 | x_i\in B_k \rangle \nonumber\\
&\qquad\qquad\qquad - \lbrack D^{(1)}(x^k)\rbrack^2\,\tau^2 \Bigr\}
\end{align}
\end{subequations}

\noindent The limit $\tau\to 0$ can be realized by a polynomial extrapolation of the values obtained for the time steps $\tau$, $2\tau$, $3\tau$, etc. Alternatively, the value for the time step $\tau$ that is given by the sampled measurement can be used as an approximation of 0th order.

A Markov process is a substantial simplification of a general stochastic process. Especially in the context of high sampling rates, there are measured time series that cannot be described by a Markov process satisfactorily \cite{ReviewFriedrichPeinke}. A possible non-Markovian generalization of the standard Langevin equation defined by Eq.~(\ref{LangevinEq}) can be made by introducing a memory kernel \cite{GLE_PRE}. However, in this work, we follow an alternative approach that has already been proposed in Ref.~\cite{Katrin2DLangevin}: By adding a second, hidden component it is possible to incorporate a correlated noise process into the standard Langevin model. This correlated noise process causes that the solution of this model is non-Markovian.

Precisely, the model that we work on is the following 2-dimensional SDE (similar models have been used in Ref.~\cite{Katrin2DLangevin, BerndOUnoise}):

\begin{subequations}
\label{modelOU}
\begin{align}
\dot{X}_t &= D^{(1)}_x(X_t) + \sqrt{D^{(2)}_x(X_t)}\, Y_t\\
\dot{Y}_t &= -\frac{1}{\theta}\,Y_t + \sqrt{\frac{1}{\theta}}\,\eta_t
\end{align}
\end{subequations}

\noindent Here, the parameter $\theta$ must be positive. $\eta_t$ represents a $\delta$-correlated Gaussian Langevin force as before.

The above equations can be regarded either as one 2-dimensional Langevin equation for $X_t$ and $Y_t$ with a vanishing diffusion function in the first component or as an ordinary differential equation (ODE) for $X_t$ which is coupled with a 1-dimensional Langevin equation for $Y_t$. We follow the latter idea and understand the process $Y_t$ as a noise function which drives the dynamics of the variable $X_t$. In this sense, the process $Y_t$ in the first equation can be compared to a Wiener process in a standard Langevin equation. Therefore, we refer to $D^{(1)}_x$ and $D^{(2)}_x$ as the drift and diffusion function of the $X$-process, respectively. Yet, as an integration of a Langevin equation, our noise process is -- in contrast to a standard Langevin model -- correlated in time. The noise process that we choose is called an Ornstein-Uhlenbeck process (OU process). It is defined by the drift function $D^{(1)}_y(y) = -\frac{1}{\theta}y$, the diffusion function $D^{(2)}_y(y)=\frac{1}{\theta}$, and the correlation length $\theta$ \cite{Risken}.

We use $X_t$ to represent the quantity whose dynamics is modeled or measured. $Y_t$ is a hidden process. When this model should be adjusted to a measured time series of $X_t$, the values of $Y_t$ are unknown. The values of $Y_t$ are only known if the model equations are used to produce a synthetic time series for $X_t$ by integration. This is a central difficulty of this model which we will discuss in detail.

With respect to the 2-dimensional process $(X_t, Y_t)$, the model fulfills the Markov property as it can be seen as a 2-dimensional Langevin equation for $(X_t, Y_t)$. Yet, for the process $X_t$ alone, the Markov property is not valid. Instead, we have

\begin{align}\label{nonMarkov}
p(x_{n+1} | x_n, x_{n-1}, ..., x_0) = p(x_{n+1} | x_n, x_{n-1}),
\end{align}

\noindent which we will explain in detail later on.

Another important property of the model is that it includes two different time scales by the different correlation lengths of the two components, which leads to a concave shape of the autocorrelation function $\text{ACF}(\tau)$ for small values of $\tau$. This phenomenon can be observed in real-world examples such as wind power modeling \cite{Pesch, LevystableSuperstatistics}.

A Langevin model exhibiting hidden values is called a \textit{partially observed diffusion}. Again, the estimation of suitable model parameters on the basis of a measured time series is an important task. Here, it would be obvious to use marginalization techniques such as the \textit{expectation-maximization algorithm} (EM algorithm) \cite{EM0, EM1, EM2} or Markov chain Monte Carlo methods (MCMC methods) \cite{MCMC0, MCMC1, MCMC2, MCMC3} in a Bayesian framework. Yet, when applied to large data sets these methods are computationally expensive. In the mentioned references, the lengths of the measured time series are in the order of magnitude of $10^2$. However, to reduce statistical errors we would like to work with time series containing a number of $10^5$, $10^6$, or $10^7$ data points. In order to be able to handle this amount of data efficiently, a version of the direct estimation method that is adopted to the non-Markovian Langevin model (Eq.~(\ref{modelOU})) would be useful.

In our understanding, a direct estimation method is characterized by three properties: It is non-parametric (with respect to $D_x^{(1)}$ and $D_x^{(2)}$), it involves a binning approach, and it relies on an analysis of increment moments. The binning approach in connection with the idea of representing the statistical properties of the data by the increment moments result in a method which is very effective when it is applied to large data sets.

In Ref.~\cite{BerndOUnoise}, an estimation method for a 2-dimensional Langevin model similar to that defined in Eq.~(\ref{modelOU}) is presented. It can be called a direct estimation method in the sense defined above and is actually computationally cheap. Yet, this procedure is based on a perturbative approach and only works under the assumption that $\theta$ is small compared to the time scale of the process $X_t$.

In the present work, we propose another direct estimation method for the non-Markovian Langevin model defined in Eq.~(\ref{modelOU}) which does not include any restriction for the time scales (cf. Sec.~\ref{sec:Estimation}). So, it can also be applied in situations in which the non-Markovian nature of the data is very pronounced. We perform the estimation by optimizing a cost function whereby the computational effort of an evaluation of this cost function is independent of the length of the measured time series. Hence, the procedure can be applied to huge data sets without any problem.

Precisely, we use the coupling of the processes $X_t$ and $Y_t$ to define one equation which incorporates the overall dynamics of our model in the sense of Eq.~(\ref{nonMarkov}). Here, different increments of the process $X_t$ are incorporated. By a reformulation of this equation, we obtain a random variable $\chi_t$ that obeys a standard normal distribution. We use its known moments as conditions for an estimation procedure. Following the idea of the direct estimation method, we sort the data into bins and obtain non-parametric estimates for the values of the functions $D_x^{(1)}$ and $D_x^{(2)}$ in these bins. As mentioned, we perform the estimation by an optimization of a suitable cost function. This is necessary, because, unlike in the direct estimation of standard Langevin equations, there is a connection between the estimates in different bins. This connection arises from the hidden component. The parameter $\theta$ appears in the equations of all bins.

We give reasons for the assumption that the solution of the estimation problem concerning the non-Markovian model is unique (cf. Sec.~\ref{sec:Optimization}). For that purpose, we consider the solution of the model in the case of linear functions $D_x^{(1)}$ and $D_x^{(2)}$, because in this particular case the solution can be calculated analytically.

The uniqueness is important for the feasibility of the optimization. It implies that there is only one global minimum and that the problem is well-posed. Nevertheless, it turns out that the cost function is difficult to handle due to different local minima. We propose two possibilities to manage the optimization efficiently in different situations (cf. Sec.~\ref{sec:Optimization}). 

Firstly, we show how to obtain a useful initial guess for the optimization procedure. This initial guess only relies on the autocorrelation function (ACF) and the variance (Var) of the measured time series. It corresponds to the linear version of the non-Markovian Langevin model (Eq.~(\ref{modelOU})). Thus, it is successful if the measured dynamics can be approximated by a linear model to a certain extend. We discuss several examples to give an idea of what that means (cf. Sec.~\ref{sec:Synth}).

Alternatively, we propose to support the optimization process by a $L^2$-regularization. Hereby, we choose the regularization parameter using the L-curve method. This ensures that the change of the cost function which is made by the regularization term does not affect the result of the estimation too much. Nonetheless, for our model, the $L^2$-regularization favors balanced values during optimization. Therefore, this approach is useful only if this has a negligible impact on the estimation problem.

In a real world example in which the true model is unknown, the decision for one of the mentioned approaches has to be made by trial and error. It is based on a comparison of the statistical properties of the measured time series and a time series generated by an estimated model. Here, we consider the onepoint pdf $p_{X_t}$, the increment pdf $p_{X_{t+\tau} - X_t}$, and the autocorrelation function $\text{ACF}$. Of course, these statistical functions do not characterize a stochastic process in a unique way, but they are important reference points.

For the purpose of validation, we apply the proposed estimation procedure on several synthetic examples (cf. Sec.~\ref{sec:Synth}). Hereby, we illustrate the abilities of the procedure in different situations. 

In sum, this paper is organized as follows. In Sec.~\ref{sec:Estimation}, we present the estimation procedure for the non-Markovian Langevin model which we introduced above. We discuss the mentioned issues concerning the optimization in Sec.~\ref{sec:Optimization}. Afterwards, we illustrate the procedure on the basis of several synthetic examples for the purpose of validation in Sec.~\ref{sec:Synth}. In Sec.~\ref{sec:Summary}, we summarize our work and give an outlook.

\section{Estimation procedure\label{sec:Estimation}}

Faced with the problem of the estimation of the non-Markovian model defined in Eq.~(\ref{modelOU}), one could try to apply the Markovian direct estimation (cf. Eq.~(\ref{directEst})) to the measured values $(x_0, x_1, x_2, ..., x_N)$. For this purpose, we perform a binning of the range $D = \lbrack x_{\min}, x_{\max} \rbrack$ in the form $D = \cup_{k=1}^{N_{\text{bins}}} B_k$, where a bin $B_k$ is centered at $x^k$ as for the 1-dimensional case. By the Euler approximation of the first component of the model, one would obtain the following equation for bin $B_k$ centered at $x^k$:

\begin{align}\label{eq7}
&\frac{1}{\tau} \langle X_{t+\tau} - X_t | X_t\in B_k \rangle \nonumber\\
&\qquad\qquad \approx D_x^{(1)}(x^k) + \sqrt{D_x^{(2)}(x^k)}\,\langle Y_t | X_t\in B_k \rangle
\end{align}

For a standard Langevin equation, the $\delta$-correlation of the noise process would cause that the mean $\langle Y_t | X_t\in B_k \rangle$ is zero for every bin $B_k$. That is why the conditional mean of the increment of $X_t$ is used to calculate an estimation of $D_x^{(1)}$ at $x^k$.

However in the case of a correlated noise process, the value of this mean depends on the location of the bin. This can be understood intuitively as follows. The noise drives the dynamics of the quantity $X$ against the force that is defined by the drift function. Hence, if $X_t$ reaches a bin $B_{k^{\ast}}$ that is far away from a fixed point at time $t^{\ast}$, the values of the noise process $Y_t$ in a small time interval before $t^{\ast}$ will be far away from zero. Due to the time correlation of the process $Y_t$, the same will be true for $Y_{t^{\ast}}$. Consequently, the mean of all values of $Y_t$ with the condition that $X_t$ is located in the bin $B_{k^{\ast}}$ is not zero. As the values of $Y_t$ are not known, Eq.~(\ref{eq7}) is not helpful in this context. The Markovian direct estimation cannot be applied to the present model.

Instead, we carry out similar estimations by using the coupling of $X_t$ and $Y_t$ to obtain one equation for the dynamics of the measured values of $X_t$.

The Euler-Maruyama approximation of the model defined by Eq.~(\ref{modelOU}) reads

\begin{subequations}
\label{ApproxModelOU}
\begin{align}
 X_{t+\tau} &= X_t + D^{(1)}_x(X_t)\,\tau + \sqrt{D^{(2)}_x(X_t)}\, Y_t\,\tau\\
 Y_{t+\tau} &= Y_t - \frac{1}{\theta}\, Y_t\,\tau + \sqrt{\frac{1}{\theta}}\sqrt{\tau}\, N_t
\end{align}
\end{subequations}

\noindent where $N_t$ is an element of a sequence of independent and identically distributed random variables obeying a standard normal distribution. Here, an approximation of higher order in $\tau$ would be possible. However, it would contain different stochastically dependent It\^{o} integrals that would make the following calculations impossible. Based on Eq.~(\ref{ApproxModelOU}), we formulate the following three relations

\begin{align}
X_{t+\tau} &= X_{t+\tau}(X_t, Y_t, \xi) \label{(i)}\\
Y_{t+\tau} &= Y_{t+\tau}(Y_t, N_t, \xi) \label{(ii)}\\
X_{t+2\tau} &= X_{t+2\tau}(X_{t+\tau}, Y_{t+\tau}, \xi), \label{(iii)}
\end{align}

\noindent where $\xi := (D_x^{(1)}, D_x^{(2)}, \theta)$ denotes the degrees of freedom of the estimated model. We combine these relations into one equation that only includes measured values of the process $X_t$ via simple reformulations as follows. From the measured values $X_t$ and $X_{t+\tau}$, the hidden value $Y_t$ can be determined via Eq.~(\ref{(i)}). Then, Eq.~(\ref{(ii)}) can be used to calculate $Y_{t+\tau}$ where the random variable $N_t$ is involved. Finally, the calculated value $Y_{t+\tau}$ and the measured value $X_{t+\tau}$ can be inserted into Eq.~(\ref{(iii)}) to gain a relation between $X_t$, $X_{t+\tau}$, and $X_{t+2\tau}$. This relation incorporates the overall dynamics of the measured process $X_t$ that is influenced by the coupled noise process $Y_t$. It makes clear that the process $X_t$ alone is non-Markovian by means of Eq.~(\ref{nonMarkov}). The relation reads

\begin{align}
X_{t+2\tau} &= X_{t+2\tau}(X_{t+\tau}, X_t, N_t, \xi) \nonumber\\
&= \varphi(X_{t+\tau}, X_t, \xi) + \psi(X_{t+\tau}, X_t, \xi)\, N_t \label{propagator}
\end{align}

\noindent where we use the abbreviations

\begin{align}
\varphi(X_{t+\tau}, X_t, \xi) &:= X_{t+\tau} \nonumber\\
&+ \left(D^{(1)}_x(X_{t+\tau}) + \sqrt{D^{(2)}_x(X_{t+\tau})}Y_t\right)\,\tau \nonumber\\
&- \sqrt{D^{(2)}_x(X_{t+\tau})}\frac{1}{\theta}\,Y_t\,\tau^2\\
\psi(X_{t+\tau}, X_t, \xi) &:= \sqrt{D^{(2)}_x(X_{t+\tau})}\sqrt{\frac{1}{\theta}}\,\tau^{3/2}
\end{align}

\noindent and the relation

\begin{align}
Y_t = \frac{X_{t+\tau} - X_t - D^{(1)}_x(X_t)\,\tau}{\sqrt{D^{(2)}_x(X_t)}\tau}.
\end{align}

From Eq.~(\ref{propagator}), we can infer that the random variable $\chi_t$ defined by

\begin{align}
\chi_{t+2\tau} := \frac{X_{t+2\tau} - \varphi(X_{t+\tau}, X_t, \xi)}{\psi(X_{t+\tau}, X_t, \xi)}
\end{align}

\noindent obeys a standard normal distribution for every value of $t$ where $\chi_t$ and $\chi_{t'}$ are independent for any different points in time $t\ne t'$.

Now, for a measured time series $(x_0, x_1,...,x_N)$ of the process $X_t$, we can calculate the corresponding sequence $(\chi_2, \chi_3,...,\chi_N)$. This sequence is a sample of a standard normal distribution. Consequently, we have

\begin{align}
\langle \chi_{i+2} \rangle = 0 \qquad \text{and} \qquad \langle \chi_{i+2}^2 \rangle = 1.
\end{align}

We use these equations as conditions for the time-independent values of $\xi$. As in the standard direct estimation, we perform the binning and calculate conditioned means to yield conditions for both $\theta$ and the drift and diffusion functions at the center of a specific bin. In this case, due to the non-Markovian nature of the model in terms of Eq.~(\ref{nonMarkov}), we need two conditions. For a bin $B_k$ centered at $x^k$, we have the equations

\begin{align}
&\langle \chi_{i+2} | x_{i+1}\in B_k, x_i\in B_k \rangle = 0 \\
\text{and}\quad &\langle \chi_{i+2}^2 | x_{i+1}\in B_k, x_i\in B_k \rangle = 1
\end{align}

\noindent as conditions for the values $D_x^{(1)}(x^k)$, $D_x^{(2)}(x^k)$ and $\theta$.

In this context, an estimated model is represented by the values

\begin{align}\label{DefXiHat}
\hat{\xi} := &(D_x^{(1)}(x^1),...,D_x^{(1)}(x^{N_{\text{bins}}}), \nonumber\\
&\quad D_x^{(2)}(x^1),...,D_x^{(2)}(x^{N_{\text{bins}}}),\theta).
\end{align}

Via the parameter $\theta$, conditions belonging to different bins are coupled. Therefore, we formulate our estimation procedure in terms of an optimization problem. To find a suitable model for a measured time series $(x_0, x_1,...,x_N)$, we minimize the \textit{cost function}

\begin{align}
\text{CF}(\hat{\xi}) := &\sum_{k=1}^{N_{\text{bins}}} \Bigl( \langle \chi_{i+2} | x_{i+1}\in B_k, x_i\in B_k \rangle - 0 \Bigr)^2 \nonumber\\
&+ \sum_{k=1}^{N_{\text{bins}}} \Bigl( \langle \chi_{i+2}^2 | x_{i+1}\in B_k, x_i\in B_k \rangle - 1 \Bigr)^2. \label{CF}
\end{align}

It is possible to reformulate the conditional means of $\chi_{i+2}$ in the following way:

\begin{align}
&\langle \chi_{i+2} | x_{i+1}\in B_k, x_i\in B_k \rangle = \nonumber\\
&\qquad\qquad\qquad E_k\Delta_k^{10} + F_k\Delta_k^{01} + G_k\Delta_k^{00}\\
&\langle \chi_{i+2}^2 | x_{i+1}\in B_k, x_i\in B_k \rangle = \nonumber\\
&\qquad\qquad\qquad E_k^2\Delta_k^{20} + 2E_kF_k\Delta_k^{11} + 2E_kG_k\Delta_k^{10} \nonumber\\
&\qquad\qquad\qquad + F_k^2\Delta_k^{02} + 2F_kG_k\Delta_k^{01} + G_k^2\Delta_k^{00}
\end{align}

\noindent Here, we use the notations

\begin{align}
\Delta_k^{mn} &:= \bigl\langle (x_{i+2}-x_{i+1})^m(x_{i+1}-x_i)^n \nonumber\\
&\qquad\qquad | x_{i+1}\in B_k, x_i\in B_k \bigr\rangle\\
E_k &:= \frac{1}{ \sqrt{D_x^{(2)}(x^k)\,\frac{1}{\theta}\,\tau^3} } \\
F_k &:= \frac{ \frac{1}{\theta}\,\tau - 1 }{ \sqrt{D_x^{(2)}(x^k)\,\frac{1}{\theta}\,\tau^3} } \\
G_k &:= -\frac{D_x^{(1)}(x^k)\sqrt{\frac{1}{\theta}\,\tau}}{\sqrt{D_x^{(2)}(x^k)}}.
\end{align}

The important advantage of this reformulation is that the quantities $\Delta_k^{mn}$, which represent the statistical properties of the measured time series containing a huge amount of data points, can be calculated once before the optimization of the cost function $\text{CF}$ is performed. In this way, the numerical effort of the optimization of the cost function is reduced considerably. Moreover, it is independent of the length of the data set.

In principle, it would be possible to include higher moments of the random variable $\chi_t$ as additional conditions into the cost function. However, this turns out to be not helpful in practice because it does not improve the result of the estimation.

The number of bins $N_{\text{bins}}$ should be chosen great enough so that the resolution of the functions $D_x^{(1)}$ and $D_x^{(2)}$ is high enough. At the same time, it should be chosen small enough so that the course of the functions is not perturbed by statistical fluctuations which appear if the amount of data belonging to a single bin is too small. Hence, the most suitable number of bins also depends on the length of the measured time series. In our examples where we have $10^6$ data points, we choose $N_{\text{bins}}=20$.

Furthermore, we modify the cost function in such a way that it returns a value that is higher than all other values, if the minimum of the function $D_x^{(2)}$ or the value of $\theta$ are negative. In this way, the constraints of the model are fulfilled by the result of the estimation.

In summary, we conduct an estimation of the model defined by Eq.~(\ref{modelOU}) by finding a suitable value of the parameter $\theta$ and suitable values of the drift and diffusion functions $D_x^{(1)}$ and $D_x^{(2)}$ at the centers of appropriate bins. The estimation procedure is realized by a minimization of the cost function $\text{CF}$ (cf. Eq.~(\ref{CF})). The numerical effort of an evaluation of this cost function is independent of the length of the data set.

\section{Technical details concerning the optimization\label{sec:Optimization}}

We expect to obtain unique results from the estimation of the model defined by Eq.~(\ref{modelOU}). We give reasons for that in Sec.~\ref{subsec:uniqueness}. Nonetheless, the optimization of the cost function $\text{CF}$ (cf. Eq.~(\ref{CF})) turns out to be difficult in practice due to different local minima. Therefore, we support the optimization process either by a suitable initial guess (cf. Sec.~\ref{subsec:iniGuess}) or by a regularization (cf. Sec.~\ref{subsec:regularization}).

\subsection{Uniqueness of the global minimum\label{subsec:uniqueness}}

For the linear model

\begin{subequations}
\label{modelLinear}
\begin{align}
\dot{X}_t &= -\frac{1}{\alpha} X_t + \sqrt{\beta}\, Y_t\\
\dot{Y}_t &= -\frac{1}{\theta}\,Y_t + \sqrt{\frac{1}{\theta}}\,\eta_t
\end{align}
\end{subequations}

\noindent ($\alpha,\beta,\theta>0$) the stationary solution of the process $X_t$ can be calculated analytically. It is (cf. Appendix~\ref{app:A})

\begin{align}\label{SolLinear0}
X_t = \sqrt{\frac{\beta}{\alpha}} \int_0^t (t-t')\, e^{-\frac{1}{\alpha}(t-t')}\, dW_{t'}
\end{align}

\noindent in the case $\alpha=\theta$ and

\begin{align}\label{SolLinear}
X_t = \frac{\alpha\sqrt{\beta\theta}}{\alpha - \theta} \int_0^t e^{-\frac{1}{\alpha}(t-t')} - e^{-\frac{1}{\theta}(t-t')} \, dW_{t'}
\end{align}

\noindent in the case $\alpha \ne \theta$. Hereby the integral is a stochastic It\^{o} integral with respect to the Wiener process $W_t$.

The corresponding variance $\text{Var}(X_t)$ and autocorrelation function $\text{ACF}(\tau)$ read (cf. Appendix~\ref{app:AB})

\begin{align}
\text{Var}(X_t) &= \frac{\alpha^2\beta}{4}\\
\text{ACF}(\tau) &= e^{-\frac{1}{\alpha}\tau} + \frac{1}{\alpha}\tau e^{-\frac{1}{\alpha}\tau} \label{ACF0}
\end{align}

\noindent in the case $\alpha=\theta$ and

\begin{align}
\text{Var}(X_t) &= \frac{\alpha^2\beta\theta}{2(\alpha+\theta)} \label{Variance}\\
\text{ACF}(\tau) &= \frac{1}{\alpha - \theta} \Bigl\lbrack \alpha e^{-\frac{1}{\alpha}\tau} - \theta e^{-\frac{1}{\theta}\tau} \Bigr\rbrack \label{ACF}
\end{align}

\noindent in the case $\alpha\ne\theta$.

With the help of these results we can investigate the uniqueness of the solution of our estimation problem.

In the linear model, the parameters $\alpha$ and $\theta$ can be interchanged without changing the solution of the model (cf. Eq.~(\ref{SolLinear0}) and Eq.~(\ref{SolLinear})). Hereby, the value of $\beta$ must be adjusted suitably in the case $\alpha\ne\theta$. Aside from this symmetry, the values of all parameters are uniquely defined by the solution $X_t$. This can be proved as follows.

The functions $e^{-\frac{1}{\alpha}\tau}$, $e^{-\frac{1}{\theta}\tau}$, and $\tau e^{-\frac{1}{\alpha}\tau}$ appearing in the ACF are linearly independent. Consequently, the values of $\alpha$ and $\theta$ can be uniquely deduced from the ACF. Besides, the cases $\alpha=\theta$ and $\alpha\ne\theta$ can be distinguished by means of the ACF. With known values of $\alpha$ and $\theta$ the variance can be used to determine the value of $\beta$.

All in all, the solution of the estimation problem concerning the linear model is unique except for the symmetry regarding the parameters $\alpha$ and $\theta$.

Using the transformation $X_t' = X_t - X^{\ast}$, these results can be generalized to a linear model in which the fixed point $X^{\ast}$ of the process $X_t$ differs from $0$. The corresponding model equations read

\begin{subequations}
\label{modelLinearShift}
\begin{align}
\dot{X}_t &= -\frac{1}{\alpha}(X_t - X^{\ast}) + \sqrt{\beta}\, Y_t\\
\dot{Y}_t &= -\frac{1}{\theta}\,Y_t + \sqrt{\frac{1}{\theta}}\,\eta_t.
\end{align}
\end{subequations}

\noindent In this case, the fixed point $X^{\ast}$ is uniquely determined by the mean $\langle X_t \rangle$ (the mean of the solution $X_t$ of the model given by Eq.~(\ref{modelLinear}) is 0).

According to our experience (cf. Sec.~\ref{sec:Synth}), it is possible to transfer the above results to the nonlinear model (Eq.~({\ref{modelOU})), whereas the symmetry regarding the slope of the function $D_x^{(1)}$ and the term $-\frac{1}{\theta}$ which occurs in the linear model is broken in the nonlinear model. Therefore, we have reasons to assume that our estimation problem is \textit{well-posed}.

An important issue in this context is that we have chosen the noise process in such a way that it includes only one free parameter. Moreover, it is chosen such that the variance of the process $Y_t$ is fixed to $\frac12$ and that $\theta$ defines the correlation length of $Y_t$ \cite{Risken}. Thus, the magnitude of the influence of the noise to the dynamics of $X$ is only determined by the function $D^{(2)}_x$. According to our experience, this is advantageous for the estimation procedure.

\subsection{Linear estimation as initial guess\label{subsec:iniGuess}}

In order to obtain an initial guess for the optimization of the cost function $\text{CF}$ (cf. Eq.~(\ref{CF})), we first conduct an estimation of the linear model (Eq.~(\ref{modelLinear})). We perform this linear estimation only based on the ACF and the variance of the measured time series as described next.

As it is often done in the context of turbulence, two time scales can be determined by the autocorrelation function $\text{ACF}(\tau)$. The \textit{correlation length} $L$ is defined by its integral

\begin{align}
L = \int_0^{\infty} \text{ACF}(\tau)\, d\tau
\end{align}

\noindent and the \textit{Taylor length} $\Lambda$ is defined by its quadratic approximation at $\tau=0$

\begin{align}
\text{ACF}(\tau) \approx 1 - \frac12 \left(\frac{\tau}{\Lambda}\right)^2.
\end{align}

\noindent There are different possibilities for a precise computation of these quantities in practice \cite{Turb:Reinke}. Yet, for the purpose of this work, it is sufficient to compute them in a very straightforward way. We calculate the correlation length $L$ by integrating the function $\text{ACF}(\tau)$ from $\tau=0$ to the point where it is zero or nearly zero for the first time. Here, we ignore the fluctuations that usually appear in a measured autocorrelation function for large values of $\tau$. For the calculation of the Taylor length $\Lambda$, we perform a quadratic fit of the function $\text{ACF}(\tau)$ in a small range close to $\tau=0$. The precise result for $\Lambda$ depends on the choice of this range. We define the most appropriate range by visual inspection such that the curvature of the ACF near $\tau=0$ is reflected sufficiently by the quadratic fit. An example is shown in Fig.~\ref{fig:ACF_quadraticFit}.

\begin{figure}
  \includegraphics[width=0.8\hsize]{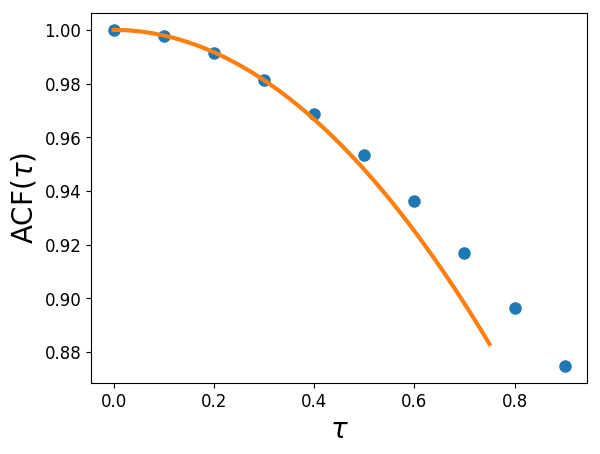}
  \caption{Autocorrelation function $\text{ACF}(\tau)$ (blue dots; belonging to the intermittent system, cf. Sec.~\ref{sec:Synth}) and quadratic fit with respect to the first four samples (orange line). The Taylor length $\Lambda$ can be derived from the second-order coefficient of the quadratic fit.\label{fig:ACF_quadraticFit}}
\end{figure}

There is a very clear connection between the length scales $L$ and $\Lambda$ of the ACF and the parameters $\alpha$ and $\theta$ of the linear model. Via Eq.~(\ref{ACF}), we obtain (cf. Appendix~\ref{app:B})

\begin{align}
L=\alpha+\theta \quad\text{and}\quad \Lambda=\sqrt{\alpha\theta}
\end{align}

\noindent in the case $\alpha\ne\theta$. Vice versa, with measured values of $L$ and $\Lambda$, we can calculate estimated values for $\alpha$ and $\theta$ by

\begin{subequations}
\label{alphathetahat}
\begin{align}
\hat{\alpha} &= \frac{L}{2} \pm \sqrt{ \left(\frac{L}{2}\right)^2 - \Lambda^2}\\
\hat{\theta} &= \frac{L}{2} \mp \sqrt{ \left(\frac{L}{2}\right)^2 - \Lambda^2}.
\end{align}
\end{subequations}

\noindent Hereby, the choice of the signs cannot be made based on the ACF. This corresponds to the symmetry of the parameters $\alpha$ and $\theta$ that we have already discussed in Sec.~\ref{subsec:uniqueness}. In our validation examples, the choice can be made by comparing $\hat{\theta}$ to the true value of $\theta$ and $-\frac{1}{\hat{\alpha}}$ to an averaged slope of the true function $D_x^{(1)}$. In a real world estimation problem, the choice can only be made based on the comparison of the statistical properties of the measured time series and a time series generated by the estimated model.

Via the estimated values $\hat{\alpha}$ and $\hat{\theta}$, Eq.~(\ref{Variance}) can be used to obtain an estimate for the parameter $\beta$:

\begin{align}\label{betahat}
\hat{\beta} = \frac{2\,L\,\text{Var}}{\hat{\alpha}^2\,\hat{\theta}}
\end{align}

The above estimates can also be applied in the case $\alpha=\theta$.

Due to inaccuracies that are involved in the determination of the length scales $L$ and $\Lambda$ in a measured ACF, the square roots in the above estimates $\hat{\alpha}$ and $\hat{\theta}$ can be problematic in the case $L/2 \approx \Lambda$. If such a problem occurs, it can be appropriate to simply use $\hat{\alpha}=\hat{\theta}=L/2$ instead of the above estimates.

Alternatively, we can employ the function of either Eq.~(\ref{ACF0}) ($\alpha=\theta$) or Eq.~(\ref{ACF}) ($\alpha\ne\theta$) to obtain estimated parameters $\hat{\alpha}$ and $\hat{\theta}$ by a fit of the measured ACF. Here again, the decision for one of the cases $\alpha=\theta$ and $\alpha\ne\theta$ as well as for the order of assignment of the fitted parameters to the estimates $\hat{\alpha}$ and $\hat{\theta}$ can only be made by trial and error.

The above described linear estimation yields a proper initial guess for the optimization of the cost function CF (Eq.~(\ref{CF})) if the measured dynamics can be approximated sufficiently by the linear model (Eq.~(\ref{modelLinear})). In  Sec.~\ref{sec:Synth} we discuss several examples to give an idea of what that means.

\subsection{Regularization\label{subsec:regularization}}

Alternatively to the linear initial guess discussed in the previous section, we can support the optimization process by a $L^2$-regularization \cite{Tikhonov}. Again, the decision in favor of this procedure can only be made by trial and error based on a comparison of the statistical properties of the measured time series and a time series generated by the estimated model.

The $L^2$-regularization is facilitated by adding the term

\begin{align}
\lambda\, \left( \sum_{k=1}^{N_{\text{bins}}} \left\lbrack D_x^{(1)}(x^k) \right\rbrack^2 + \sum_{k=1}^{N_{\text{bins}}} \left\lbrack D_x^{(2)}(x^k) \right\rbrack^2 + 2N_{\text{bins}}\,\left(\frac{1}{\theta}\right)^2 \right)
\end{align}

\noindent to the cost function $\text{CF}$ (cf. Eq.~(\ref{DefXiHat}) and (\ref{CF})). The factor $2N_{\text{bins}}$ ensures that the regularization reaches the noise process to the same extend as the first component of the model.

We determine the regularization parameter $\lambda$ using the L-curve method \cite{Tikhonov}. This method ensures that the cost function is only modified as much as it is necessary to obtain a reliable optimization procedure. Yet, the value of $\lambda$ is not always unique. Sometimes it can be helpful to try different candidates.

Due to the connection of the two components of the model via the correlation length $L$ ($L=\alpha + \theta$ in the linear case), the $L^2$-regularization favors balanced values of the estimated parameters. Hence, this approach is only successful, if this causes a negligible change of the optimization problem.

It is advantageous to perform the optimization step by step. We begin with an estimation of a linear model as it is defined by Eq.~(\ref{modelLinear}) or Eq.~(\ref{modelLinearShift}). This can be done with arbitrary initial guesses for the parameters $\alpha$, $X^{\ast}$, $\beta$, and $\theta$. Afterwards, we perform the non-parametric optimization by using the result of the first optimization as an initial guess. Here, we choose $N_{\text{bins}}=10$ at first and raise this value to 20 in a third step. In this way, we increase the number of free parameters successively from 4 (or 3, respectively) to 21 and 41. In every step, we carry out the estimation by means of an optimization of the cost function CF (Eq.~(\ref{CF})) with the added regularization term, whereby we choose a new regularization parameter $\lambda$ using the L-curve method in each of the three steps. This successive approach leads to more promising results.

In our experience, a regularization is not needed if a reliable initial guess can be made. Thus, it suffices to use either the regularization or the initial guess for the optimization process.

\section{Validation examples\label{sec:Synth}}

In this section, we apply our estimation procedure to different test examples for the purpose of illustration and validation.

In every example, we choose a model (by defining the drift and diffusion functions of the first component and the parameter $\theta$) and generate a time series containing $10^6$ data points by numerical integration (Euler-Maruyama) with time step $\tau = 0.1$. Next, we apply our estimation procedure (as described in Sec.~\ref{sec:Estimation} and \ref{sec:Optimization}) to this synthetic time series and compare the estimated model to the true one. Here, the comparison can be done in terms of the values of $D_x^{(1)}$, $D_x^{(2)}$, and $\theta$.

In the context of a real world example, the quality of an estimated model can only be evaluated by a comparison of the statistical properties of the measured time series and a time series generated by an integration of the model. The statistical properties of a stochastic process $X_t$ can, among other quantities, be represented by the onepoint pdf $p_{X_t}$, the increment pdf $p_{X_{t+\tau} - X_t}$, and the autocorrelation function $\text{ACF}$. Even though these statistical functions do not characterize a stochastic process in a unique way, they are important indicators.

For our test examples, we compare the statistical quantities as well. This gives a more complete view on the quality of the estimated models. Moreover, if a trial-and-error choice (number of bins, initial guess or regularization, signs in the estimates $\hat{\alpha}$ and $\hat{\theta}$, regularization parameter, ...) is not clear on the basis of the results for $D_x^{(1)}$, $D_x^{(2)}$, and $\theta$, we make the choice based on the comparison of the statistical quantities. Hereby, we show the increment probability distribution in a semi-logarithmic plot, because in this way the shape of the tails is more visible. In many applications such as turbulence or wind power modeling, this is an important issue \cite{WindPowerTurb}.

For an integration of an estimated model, we interpolate the estimated values of the drift and diffusion functions $D_x^{(1)}$ and $D_x^{(2)}$ linearly and extrapolate them constantly by the leftmost or rightmost estimated value, respectively.

We start with a simple linear model and proceed with more complicated models raising the complexity step by step to discuss the abilities of our procedure.

For all examples, we present the results of the linear estimations in Table~\ref{table}, the results of the nonlinear estimations in Fig.~\ref{fig:D12xy}, and the results in terms of the statistical quantities in Fig.~\ref{fig:Statistics}.

All estimated models include only small inaccuracies in the values of $D_x^{(1)}$, $D_x^{(2)}$, and $\theta$ as well as in the statistical characteristics. Larger deviations occur only in the values of $D_x^{(1)}$ and $D_x^{(2)}$ in the leftmost and rightmost regions of their domain. These are statistical errors that arise due to the fact that the outer bins contain less amounts of data (according to the onepoint pdf $p_{X_t}$).

\begin{table}
\caption{Estimated values for the correlation length ($L$), Taylor length ($\Lambda$), and variance (Var) of the time series belonging to the examples discussed in Sec.~\ref{sec:Synth}. Based on these quantities estimated values for the parameters $\alpha$, $\beta$, and $\theta$ of the linear approximation of the system (cf. Eq.~(\ref{modelLinear})) are calculated, which are used as an initial guess for the estimation of the general model (cf. Eq.~(\ref{modelOU})), as described in Sec.~\ref{sec:Optimization}.  \label{table}}
\begin{ruledtabular}
\begin{tabular}{l||c|c|c||c|c|c}
system & $L$ & $\Lambda$ & Var & $\hat{\alpha}$ & $\hat{\theta}$ & $\hat{\beta}$ \\
\hline\hline
OU & 1.93 & 1.00 & 0.27 & 0.96 & 0.96 & 1.14 \\
multipl. noise & 5.44 & 1.58 & 0.07 & 0.51 & 4.93 & 0.59 \\
nonlinear & 1.05 & 0.31 & 0.03 & 0.95 & 0.10 & 0.70 \\
intermittent & 3.04 & 1.51 & 0.02 & 1.35 & 1.69 & 0.04 \\
bistable & 6.81 & 1.90 & 0.99 & 6.23 & 0.58 & 0.60 \\
\end{tabular}
\end{ruledtabular}
\end{table}

\subsection{Ornstein-Uhlenbeck system\label{subsec:OUsystem}}

First of all, we choose an Ornstein-Uhlenbeck (OU) model, i.e., a model with linear drift and constant diffusion functions, given by the following equation.

\begin{subequations}
\label{eq:OU}
\begin{align}
\dot{X}_t &= -X_t + Y_t\\
\dot{Y}_t &= -Y_t + \eta_t
\end{align}
\end{subequations}

In this case, we have $\alpha=\theta$ and $L\approx\Lambda$. Due to small inaccuracies in the calculation of $L$ and $\Lambda$, the square root in the estimates $\hat{\alpha}$ and $\hat{\theta}$ (cf. Eq.~(\ref{alphathetahat})) proves problematic. Therefore, we use the function in Eq.~(\ref{ACF0}) to obtain the estimates $\hat{\alpha}$ and $\hat{\theta}$ by a fit of the measured ACF. Via Eq.~(\ref{betahat}) and the measured variance, we calculate the estimate $\hat{\beta}$ (cf. Table~\ref{table} for all numerical values).

Next, we use these results concerning the linear model (Eq.~(\ref{modelLinear})) as an initial guess for the optimization of the cost function CF (Eq.~(\ref{CF})) which yields the estimate of the nonlinear model (Eq.~(\ref{modelOU})) (cf. Fig.~\ref{fig:D12xy} and Fig.~\ref{fig:Statistics}).

\subsection{Multiplicative noise system\label{subsec:MultNoise}}

Now, we discuss a model exhibiting multiplicative noise, i.e., a nonlinear diffusion function in the first component. We choose an asymmetrical diffusion function which leads to a non-negative skew of the distribution of $X_t$. In this example, the correlation length of the noise process ($\theta=5$) equals the correlation length of the measured process $X_t$ ($L=5.44$). Hence, this estimation problem could not be solved by the method discussed in Ref.~\cite{BerndOUnoise}.

\begin{subequations}
\label{eq:MultNoise}
\begin{align}
\dot{X}_t &= -2\, X_t + \sqrt{e^{-0.5\, (X_t-1)^2}}\, Y_t\\
\dot{Y}_t &= -0.2\,Y_t + \sqrt{0.2}\,\eta_t
\end{align}
\end{subequations}

Here, we use the estimates $\hat{\alpha}$ and $\hat{\theta}$ as defined in Eq.~(\ref{alphathetahat}) and the estimate $\hat{\beta}$ as defined in Eq.~(\ref{betahat}) (cf. Table~\ref{table} for all numerical values).

Due to the asymmetric diffusion function, the mean of the measured values $\langle X_t \rangle = 0.03$ differs significantly from zero. Therefore, we use the shifted linear model (Eq.~(\ref{modelLinearShift})) as an initial guess for the optimization of the cost function CF (Eq.~(\ref{CF})) which yields the estimate of the nonlinear model (Eq.~(\ref{modelOU})) (cf. Fig.~\ref{fig:D12xy} and Fig.~\ref{fig:Statistics}).

\subsection{Nonlinear system\label{subsec:nonlinear}}

Now, we test the performance of our procedure in the context of a model exhibiting nonlinear drift and diffusion functions in the first component. Here, we choose a correlation length of the noise process ($\theta=0.1$) that is very small compared to the correlation length of the measured process $X_t$ ($L=1.05$).

\begin{subequations}
\label{eq:nonlinear}
\begin{align}
\dot{X}_t &= -X_t - X_t^3 + \sqrt{0.2\,(X_t + 2)^2}\, Y_t\\
\dot{Y}_t &= -10\,Y_t + \sqrt{10}\,\eta_t
\end{align}
\end{subequations}

Due to the small correlation length of the noise process, we need a high resolution of the ACF to be able to calculate the Taylor length $\Lambda$. Therefore, we use $\tau=0.01$ as step size of the numerical integration in this example. For a time series with step size $\tau=0.1$, a modeling with a standard Langevin equation and $\delta$-correlated noise would be more appropriate.

We use the estimates $\hat{\alpha}$ and $\hat{\theta}$ as defined in Eq.~(\ref{alphathetahat}) and $\hat{\beta}$ as defined in Eq.~(\ref{betahat}) (cf. Table~\ref{table} for all numerical values).

Afterwards, we use these results for the linear model (Eq.~(\ref{modelLinear})) as an initial guess for the optimization of the cost function CF (Eq.~(\ref{CF})) which yields the estimate of the nonlinear model (Eq.~(\ref{modelOU})) (cf. Fig.~\ref{fig:D12xy} and Fig.~\ref{fig:Statistics}).

\subsection{Intermittent system\label{subsec:intermittent}}

Here, we present another nonlinear example. The shape of the drift and diffusion functions and the correlation length of the noise process $\theta=1.43$ cause an intermittent behavior of the process $X_t$ which can be seen in the heavy tails of the increment distribution (cf. Fig.~\ref{fig:Statistics}). The term $- 0.1\,X_t^9$ in the drift function prevents the time series from diverging to infinity.

\begin{subequations}
\label{eq:intermittent}
\begin{align}
\dot{X}_t &= -0.25\,\arctan(4\,X_t) - 0.1\,X_t^9 \nonumber\\
&\quad\,+ \sqrt{0.03 + 0.2\,X_t^2}\, Y_t\\
\dot{Y}_t &= -0.7\,Y_t + \sqrt{0.7}\,\eta_t
\end{align}
\end{subequations}

A reliable estimation of the values of the drift and diffusion functions $D^{(1)}_x$ and $D^{(2)}_x$ is not feasible if a bin contains only a few data points. Therefore, in this intermittent example, we estimate $D_x^{(1)}$ and $D_x^{(2)}$ only in the interval $\lbrack -0.8, 0.8 \rbrack$.

The question how the estimated drift and diffusion functions should be extrapolated beyond this interval is a different matter. Here, we simply add the known term $- 0.1\,X_t^9$ when doing the integration of the estimated model.

Aside from that, we perform the estimation in the same way as in the previous example obtaining the results shown in Fig.~\ref{fig:D12xy} and Fig.~\ref{fig:Statistics}.

\subsection{Bistable system\label{subsec:bistable}}

In this example, we demonstrate that our estimation procedure can also handle systems that exhibit more than one stable fixed point. We take the following system into account whose dynamics is influenced by a double well potential.

\begin{subequations}
\label{eq:bistable}
\begin{align}
\dot{X}_t &= X_t - X_t^3 + \sqrt{0.9 + 0.2\, X_t^2}\, Y_t\\
\dot{Y}_t &= -\,Y_t + \eta_t
\end{align}
\end{subequations}

By trial-and-error, we decide to apply the $L^2$-regularization together with an arbitrary initial guess in this example. We perform the optimization in three successive steps as described in Sec.~\ref{subsec:regularization}. We choose the regularization parameters $\lambda_1=0.1$, $\lambda_2=10^{-6}$, and $\lambda_3=0$ by means of the L-curve method. The fact that the regularization parameter decreases from 0.1 to 0 indicates that the iterative optimization is an effective approach.

The estimation results are shown in Fig.~\ref{fig:D12xy} and Fig.~\ref{fig:Statistics}.

\begin{figure*}
  \includegraphics[width=0.33\hsize]{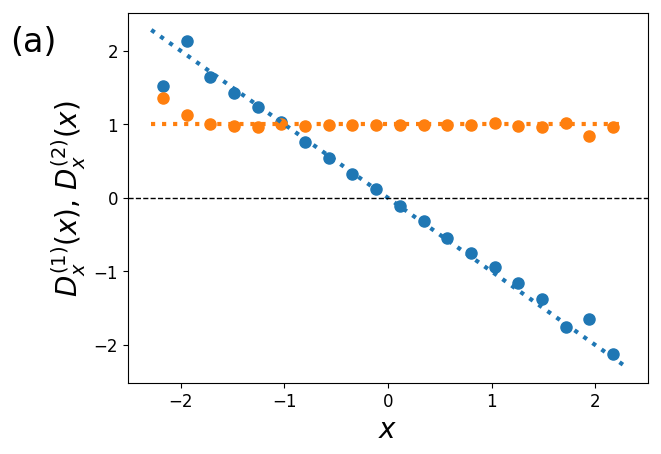}
  \includegraphics[width=0.3\hsize]{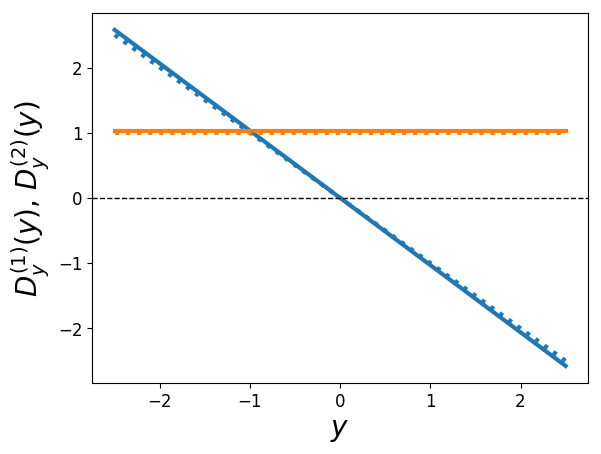}\\
  \includegraphics[width=0.33\hsize]{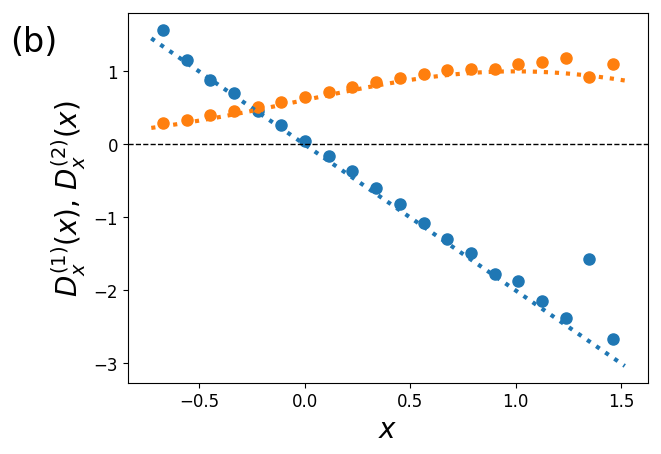}
  \includegraphics[width=0.3\hsize]{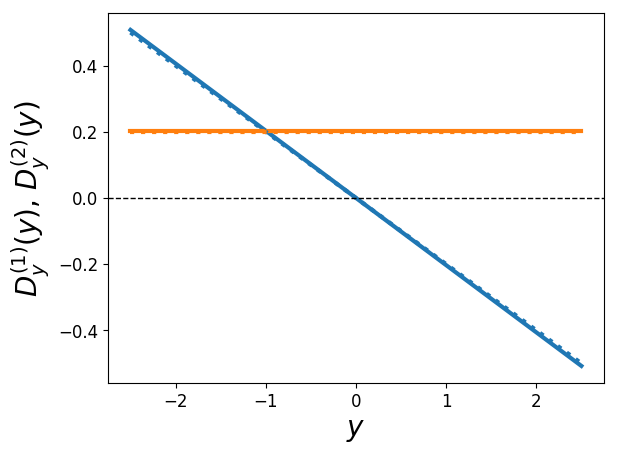}\\
  \includegraphics[width=0.33\hsize]{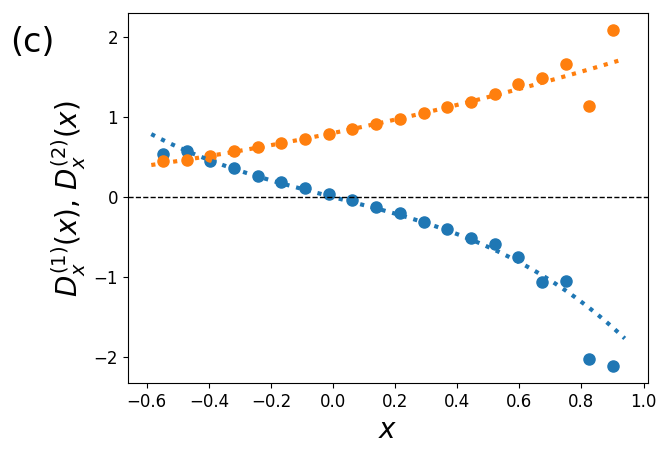}
  \includegraphics[width=0.3\hsize]{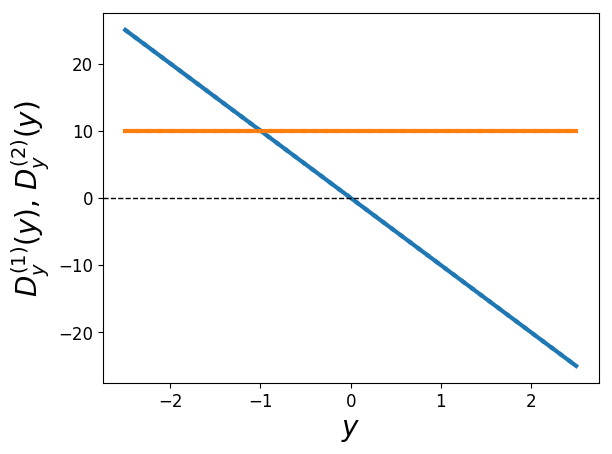}\\
  \includegraphics[width=0.33\hsize]{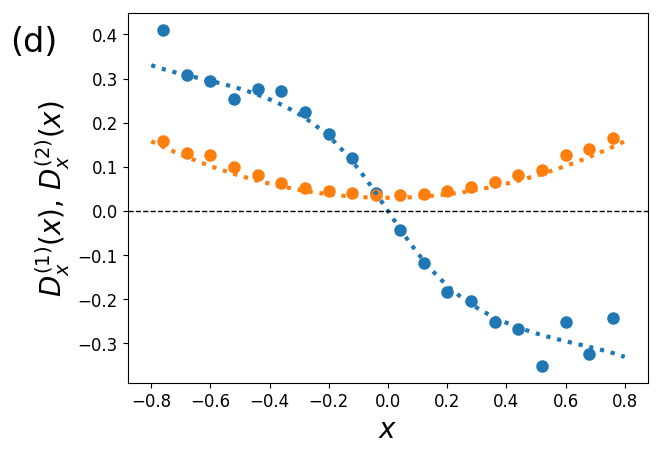}
  \includegraphics[width=0.3\hsize]{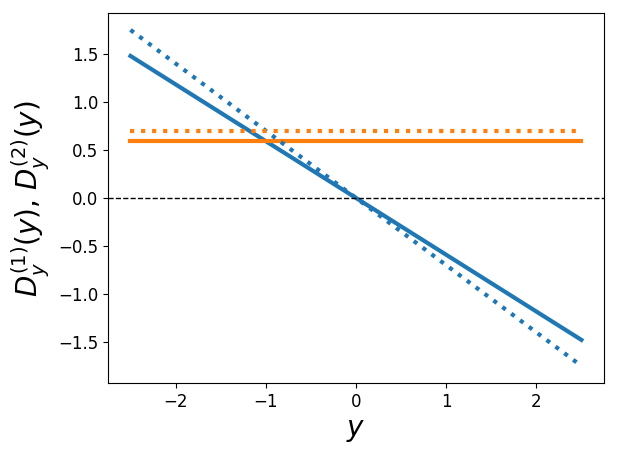}\\
  \includegraphics[width=0.33\hsize]{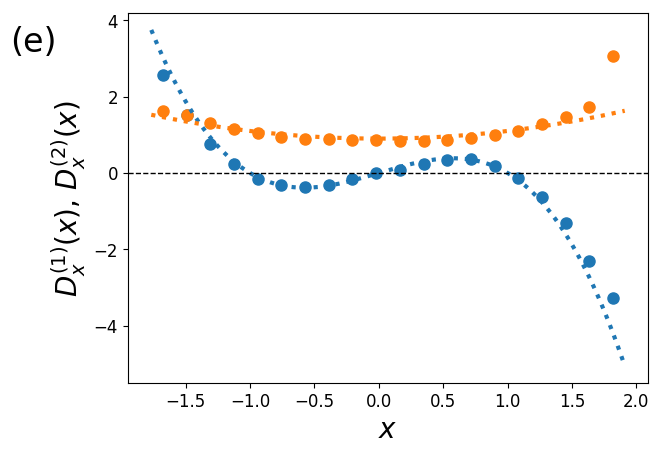}
  \includegraphics[width=0.3\hsize]{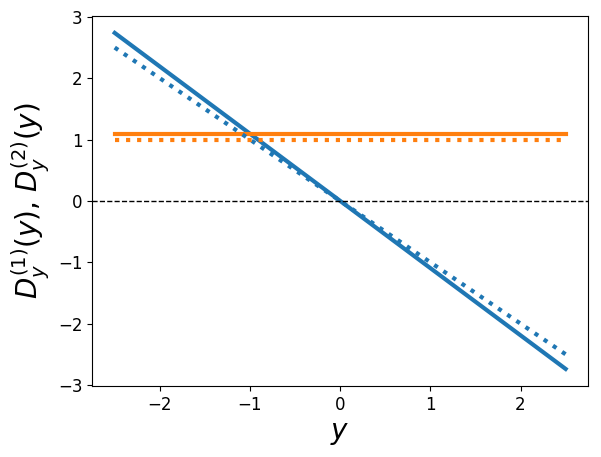}
  
  \caption{Drift (blue) and diffusion (orange) functions for $X$ (left) and $Y$ (right), respectively. The dashed lines represent the true functions. The separated dots in the left figures represent the estimated values in the respective bins. The solid lines in the right figures represent the estimated functions $D^{(1)}_y(y) = -\frac{1}{\hat{\theta}}y$ and $D^{(2)}_y(y) = \frac{1}{\hat{\theta}}$. From top to bottom: (a) OU ($\hat{\theta}=0.97$), (b) multiplicative noise ($\hat{\theta}=4.92$), (c) nonlinear ($\hat{\theta}=0.10$), (d) intermittent ($\hat{\theta}=1.69$), and (e) bistable ($\hat{\theta}=0.91$) system. \label{fig:D12xy}}
\end{figure*}

\begin{figure*}
  \includegraphics[width=0.33\hsize]{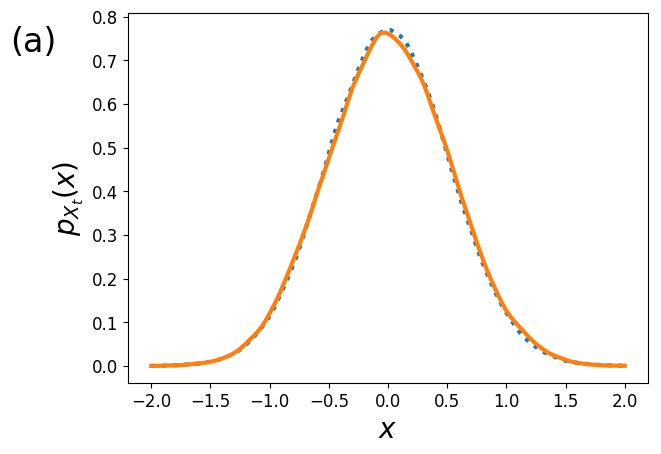}
  \includegraphics[width=0.3\hsize]{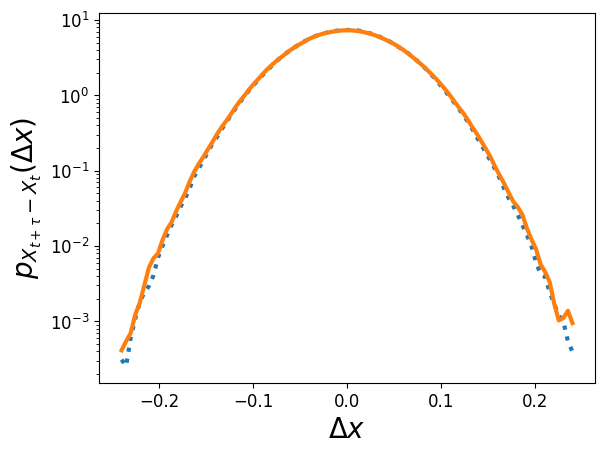}
  \includegraphics[width=0.3\hsize]{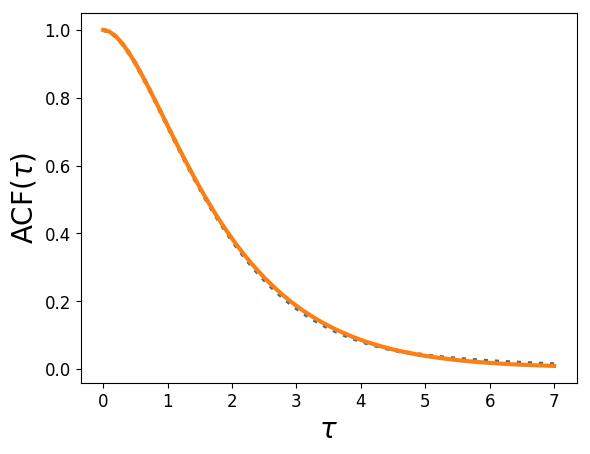}\\
 
  \includegraphics[width=0.33\hsize]{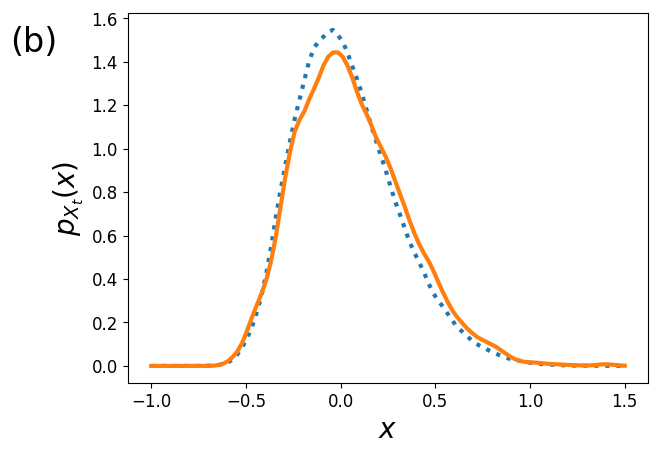}
  \includegraphics[width=0.3\hsize]{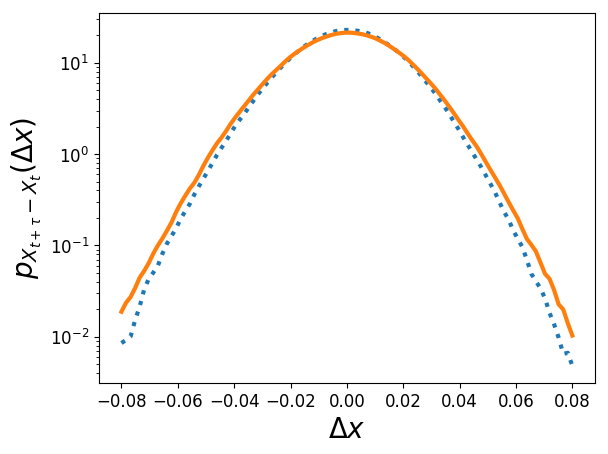}
  \includegraphics[width=0.3\hsize]{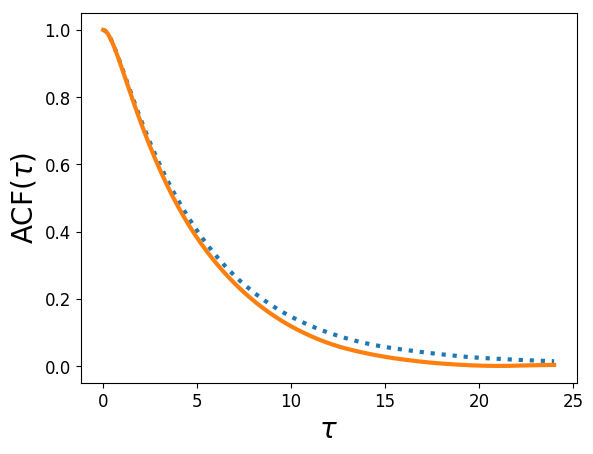}\\
  
  \includegraphics[width=0.33\hsize]{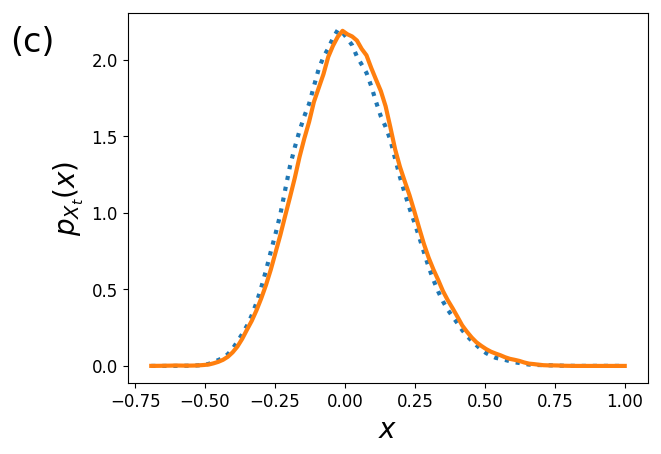}
  \includegraphics[width=0.3\hsize]{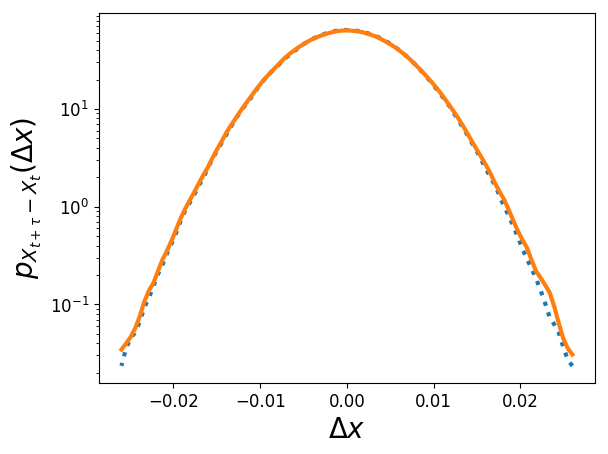}
  \includegraphics[width=0.3\hsize]{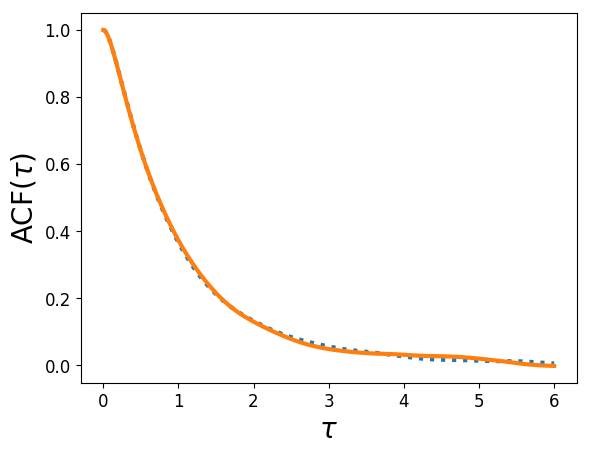}\\
  
  \includegraphics[width=0.33\hsize]{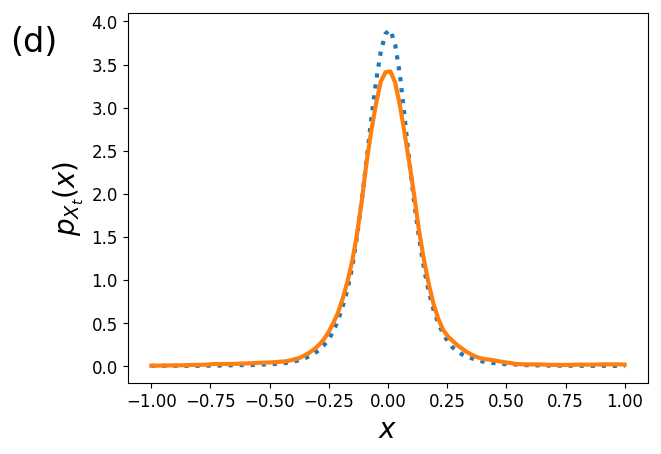}
  \includegraphics[width=0.3\hsize]{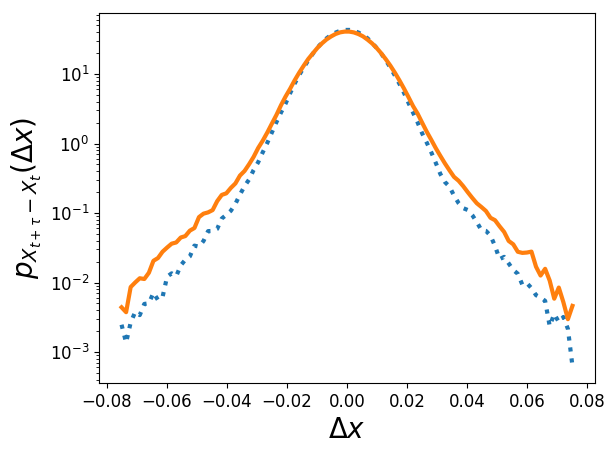}
  \includegraphics[width=0.3\hsize]{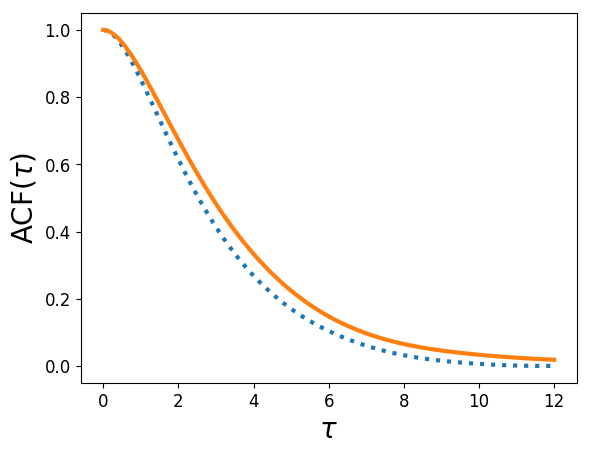}\\
  
  \includegraphics[width=0.33\hsize]{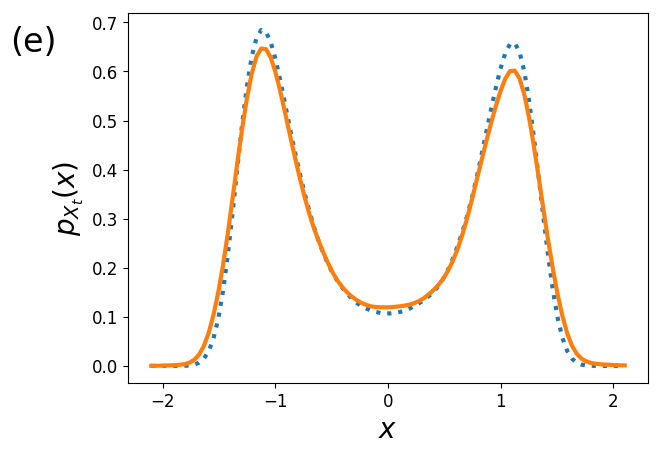}
  \includegraphics[width=0.3\hsize]{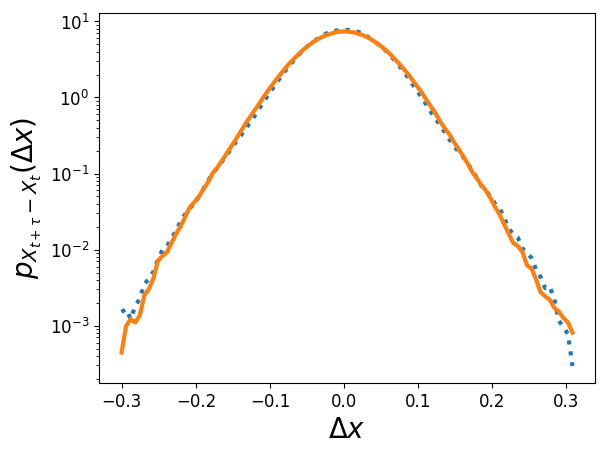}
  \includegraphics[width=0.3\hsize]{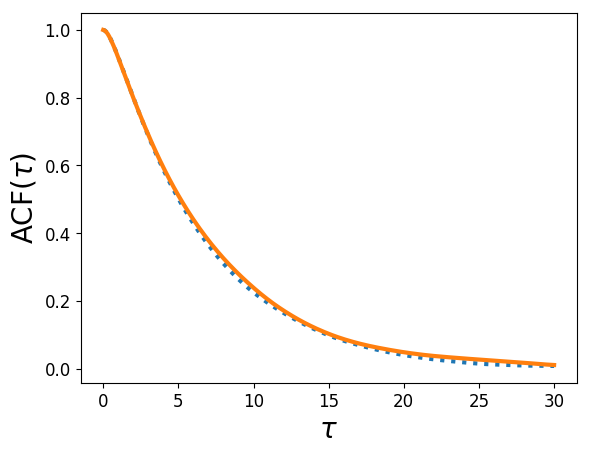}
  
  \caption{Onepoint pdf $p_{X_t}$ (left), increment pdf $p_{X_{t+\tau} - X_t}$ (middle, logarithmic plot), and autocorrelation function $\text{ACF}$ (right) of a time series of the original (dashed, blue) and estimated (solid, orange) model equations, respectively. From top to bottom: (a) OU, (b) multiplicative noise, (c) nonlinear, (d) intermittent, and (e) bistable system. Pdfs are computed by Gaussian kernel density estimation.\label{fig:Statistics}}
\end{figure*}

\section{Conclusion and Outlook\label{sec:Summary}}

In the present work, we have proposed a non-parametric estimation procedure for a non-Markovian Langevin model (cf. Eq.~(\ref{modelOU})). In this model, the dynamics of a 1-dimensional quantity is driven by a hidden Ornstein-Uhlenbeck noise process. The estimation is done by an optimization of a suitable cost function which is defined on the basis of specific means of the increment process. These means are conditioned on a suitable binning of the measured values which makes the non-parametric estimation possible. The definition of the cost function incorporates the non-Markovian nature of the model and solves the difficulty of the unknown noise values.

The numerical effort of an evaluation of the cost function is independent of the length of the measured time series. Thus, the proposed method is efficient also for large data sets. In our examples, each estimation could be realized in a few seconds on a standard PC.

Besides, we have developed a possibility to estimate the linear model (cf. Eq.~(\ref{modelLinear}) or (\ref{modelLinearShift})) which is only based on the ACF and the variance of the measured time series. It can be used to generate an initial guess for the estimation of the nonlinear model (cf. Eq.~(\ref{modelOU})). Via this initial guess or a $L^2$-regularization, we realized the optimization of the cost function in an efficient way.

As a validation, we have illustrated the abilities of the procedure in different synthetic examples. Here, we have seen only small inaccuracies in the estimated models as well as in the statistical properties of the generated time series. Solely in the outer regions of their domain, the estimated drift and diffusion functions $D_x^{(1)}$ and $D_x^{(2)}$ show remarkable deviations. As mentioned, the reason for these statistical errors is that the outer bins contain less amounts of data.

In our examples we defined the edges of the bins in a linear way. Alternatively, one could define them in such a way that all bins contain the same amount of data points. This would reduce the fluctuations of the values in the outer bins that we observed in our results. On the other hand, as a trade-off, the spacing of the outer estimated function values would be larger, resulting in lower resolution of the drift and diffusion functions in these areas.

The calculation of the initial guess as proposed requires that the measured dynamics can be approximated by the linear model to a certain extend. The regularization approach in turn is restricted to examples in which balanced values of the drift and diffusion functions and the parameter $\theta$ are reasonable. In other cases, the proposed estimation technique could produce less persuasive results. Then, a maximum likelihood estimation \cite{MLE_Prop_inPrep} could be used to improve the results. With the result of the present method as an initial guess, the computation time of the MLE could be significantly lower.

Yet, as we have seen in the examples, the restriction caused by the linear initial guess is not a significant one. Even in the intermittent example, the linear initial guess was successful. Only in the bistable example, in which the drift function exhibits several zeros, we had to rely on the regularization approach. 

The final estimated values of $\theta$ are very similar to the values that we obtained initially from the linear estimation (except for the bistable example). This fact emphasizes the effectiveness of the linear estimation that is only based on the measured ACF and variance.

The non-Markovian model that we used (Eq.~(\ref{modelOU})) can be generalized by using a nonlinear noise process. In this situation a 2-dimensional binning with respect to both the measured values of $X_t$ and the estimated values of $Y_t$ has to be performed. However, the optimization of the corresponding cost function turns out to be not feasible with the procedures discussed in this work. A possible reason for that is that the connection between the different bins is less strong than it is in our model where the parameter $\theta$ occurs in every bin. Furthermore, the 2-dimensional binning effects a higher numerical cost of the evaluation of the cost function.

We expect that the non-Markovian model that incorporates an OU noise process can be applied to many different systems. The examples that we presented in Sec.~\ref{sec:Synth} show a variety of different properties ranging from Gaussian over skewed to heavy-tailed and bimodal distributions. Moreover, an important point is that the non-Markovian model can realize an autocorrelation function $\text{ACF}(\tau)$ that is concave for small values of $\tau$. This effect is caused by the two different correlation length scales that are incorporated in the model and can be observed in real-world examples such as wind power modeling \cite{Pesch, LevystableSuperstatistics}.

In Ref.~\cite{Katrin2DLangevin}, this model is investigated qualitatively in the context of wind power data with promising results. With the estimation method that we propose here, a quantitative analysis is possible and feasible with a low numerical effort.

In the context of real world examples, the time step $\tau$ can be chosen arbitrarily. Choosing different values $\tau'=c\tau$ only leads to a scaling of the estimated values by the factor $c^{-1}$ (for $D^{(1)}_x$ and $\theta$) or $c^{-2}$ (for $D^{(2)}_x$), respectively (cf. Eq.~(\ref{numericalIntegration})).

\appendix
\section{Analytic solution of the linear model\label{app:A}}

Here, we calculate the stationary solution of the SDE model defined by the equations ($\alpha,\beta,\theta>0$)

\begin{subequations}
\label{modelLinear:app}
\begin{align}
\dot{X}^{(1)}_t &= -\frac{1}{\alpha} X^{(1)}_t + \sqrt{\beta}\, X^{(2)}_t\\
\dot{X}^{(2)}_t &= -\frac{1}{\theta}\,X^{(2)}_t + \sqrt{\frac{1}{\theta}}\,\eta_t
\end{align}
\end{subequations}

\noindent in an analytic way. In general, the solution of the initial value problem defined by the Langevin equation

\begin{align}
dX_t = -B X_t\, dt + \sigma\, dW_t,
\end{align}

\noindent and the initial value $X_0 = x_0$ is given by \cite{Risken, KloedenPlaten}

\begin{align}
X_t = e^{-Bt}x_0 + \int_0^t e^{-B(t-t')}\sigma\, dW_{t'}.
\end{align}

In our model, we have $X_t = (X^{(1)}_t, X^{(2)}_t)^\top$, $W_t = (W^{(1)}_t, W^{(2)}_t)^\top$, $B = \left( \begin{matrix} \frac{1}{\alpha} & -\sqrt{\beta}\\ 0 & \frac{1}{\theta} \end{matrix} \right)$, and $\sigma = \left( \begin{matrix} 0 & 0\\ 0 & \sqrt{\frac{1}{\theta}} \end{matrix} \right)$. As we are interested in the stationary solution that emerges for large values of $t$, we end up with $X_t^{\text{st}} = \int_0^t e^{-B(t-t')}\sigma\, dW_{t'}$ and need to evaluate the matrix exponential. Here, we have to distinguish two cases: $\alpha\ne\theta$ and $\alpha=\theta$.

In the case $\alpha\ne\theta$, we evaluate the matrix exponential via the diagonalization of $B$,

\begin{align}
B = T^{-1}DT,
\end{align}

\noindent where $T=\left( \begin{matrix} 1 & \sqrt{\beta}\frac{\alpha\theta}{\alpha-\theta}\\ 0 & 1 \end{matrix} \right)$ and $D=\left( \begin{matrix} \frac{1}{\alpha} & 0\\ 0 & \frac{1}{\theta} \end{matrix} \right)$:

\begin{align}
e^{-B(t-t')} = T^{-1}e^{-D(t-t')}T 
\end{align}

\noindent We obtain

\begin{align}\label{SolLinear:app}
X^{\text{st},(1)}_t = \frac{\alpha\sqrt{\beta\theta}}{\alpha - \theta} \int_0^t e^{-\frac{1}{\alpha}(t-t')} - e^{-\frac{1}{\theta}(t-t')} \, dW_{t'}^{(2)}
\end{align}

\noindent for the first component of the solution and

\begin{align}
X^{\text{st},(2)}_t = \frac{1}{\sqrt{\theta}} \int_0^t e^{-\frac{1}{\theta}(t-t')} \, dW_{t'}^{(2)}
\end{align}

\noindent for the second component.

In the case $\alpha=\theta$, we calculate the matrix exponential via the Jordan-Chevalley decomposition of $B$,

\begin{align}
B = A + N,
\end{align}

\noindent where $A=\left( \begin{matrix} \frac{1}{\alpha} & 0\\ 0 & \frac{1}{\alpha} \end{matrix} \right)$ is diagonal, $N=\left( \begin{matrix} 0 & -\sqrt{\beta}\\ 0 & 0 \end{matrix} \right)$ is nilpotent, and $AN=NA$:

\begin{align}
e^{-B(t-t')} = e^{-A(t-t')} e^{-N(t-t')}
\end{align}

\noindent We obtain

\begin{align}\label{SolLinear0:app}
X^{\text{st},(1)}_t = \sqrt{\frac{\beta}{\alpha}} \int_0^t (t-t')\, e^{-\frac{1}{\alpha}(t-t')}\, dW_{t'}
\end{align}

\noindent for the first component of the solution. The second component is the same as in the latter case.

\section{Variance and ACF of the linear model\label{app:AB}}

At first, we calculate the autocovariance function $\text{Cov}(X_t,X_s)$ for the first component of the stationary solution of the linear model in the case $\alpha\ne\theta$ (cf. Appendix~\ref{app:A}; for the sake of readability, we write $X_t$ instead of $X_t^{(1)}$ and $W_t$ instead of $W_t^{(1)}$):

\begin{align}
\text{Cov}(X_t, X_s) = \bigl\langle (X_t - \langle X_t\rangle)(X_s - \langle X_s\rangle) \bigr\rangle
\end{align}

Due to $\langle X_t \rangle=0$, we have

\begin{align}
&\text{Cov}(X_t, X_s) = \bigl\langle X_t X_s \bigr\rangle \nonumber\\
&\quad= \frac{\alpha^2\beta\theta}{(\alpha - \theta)^2} \bigl\langle (I_{0,t}^{t,\alpha} - I_{0,t}^{t,\theta}) (I_{0,s}^{s,\alpha} - I_{0,s}^{s,\theta}) \bigr\rangle \nonumber\\
&\quad= \frac{\alpha^2\beta\theta}{(\alpha - \theta)^2} \Bigl( \bigl\langle I_{0,t}^{t,\alpha}I_{0,s}^{s,\alpha}\bigr\rangle - \bigl\langle I_{0,t}^{t,\alpha}I_{0,s}^{s,\theta}\bigr\rangle \nonumber\\
&\quad\qquad\qquad\qquad - \bigl\langle I_{0,t}^{t,\theta}I_{0,s}^{s,\alpha}\bigr\rangle + \bigl\langle I_{0,t}^{t,\theta}I_{0,s}^{s,\theta} \bigr\rangle \Bigr),
\end{align}

\noindent where we define

\begin{align}
I_{a,b}^{c,d} := \int_a^b e^{-\frac{1}{d}(c-t')}\, dW_{t'}.
\end{align}

Representative for the four summands, we calculate $\langle I_{0,t}^{t,\alpha}I_{0,s}^{s,\theta}\rangle$. Hereby we use the relation $I_{a,b}^{c,d} = I_{a,b'}^{c,d} + I_{b',b}^{c,d}$ for $a<b'<b$. Further, $I_{a,b}^{c,d}$ and $I_{a',b'}^{c',d'}$ are stochastically independent, if $a<b\le a'<b'$, and $\langle I_{a,b}^{c,d} \rangle = 0$. Without any loss of generality, we assume $t>s$. All in all, we obtain

\begin{align}
\langle I_{0,t}^{t,\alpha}I_{0,s}^{s,\theta}\rangle &= \bigl\langle (I_{0,s}^{t,\alpha} + I_{s,t}^{t,\alpha})I_{0,s}^{s,\theta}\bigr\rangle \nonumber\\
&= \bigl\langle I_{0,s}^{t,\alpha}I_{0,s}^{s,\theta} \bigr\rangle + \bigl\langle I_{s,t}^{t,\alpha}\bigr\rangle \bigl\langle I_{0,s}^{s,\theta}\bigr\rangle = \bigl\langle I_{0,s}^{t,\alpha}I_{0,s}^{s,\theta} \bigr\rangle.
\end{align}

\noindent Now, we replace the It\^{o} integrals by approximating sums. With a partition $0=t_0<t_1<...<t_N=s$, we have

\begin{align}
\bigl\langle I_{0,s}^{t,\alpha}I_{0,s}^{s,\theta} \bigr\rangle = \lim_{N\to\infty} \Biggl\langle &\left(\sum_{i=0}^{N-1}e^{-\frac{1}{\alpha}(t-t_i)}(W_{t_{i+1}}-W_{t_i})\right) \nonumber\\
&\left(\sum_{i=0}^{N-1}e^{-\frac{1}{\theta}(s-t_i)}(W_{t_{i+1}}-W_{t_i})\right)  \Biggr\rangle.
\end{align}

\noindent $(W_{t_{i+1}}-W_{t_i})$ and $(W_{t_{j+1}}-W_{t_j})$ are stochastically independent, if $i\ne j$. Further, $\langle (W_{t_{i+1}}-W_{t_i}) \rangle = 0$. Consequently, when expanding the sums, the mixed terms vanish and we end up with

\begin{align}
&\bigl\langle I_{0,s}^{t,\alpha}I_{0,s}^{s,\theta} \bigr\rangle \nonumber\\
&= \lim_{N\to\infty} \Biggl\langle \sum_{i=0}^{N-1}e^{-\frac{1}{\alpha}(t-t_i)}e^{-\frac{1}{\theta}(s-t_i)}(W_{t_{i+1}}-W_{t_i})^2 \Biggr\rangle \nonumber\\
&=\lim_{N\to\infty} \sum_{i=0}^{N-1}e^{-\frac{1}{\alpha}(t-t_i)}e^{-\frac{1}{\theta}(s-t_i)}(t_{i+1} - t_i).
\end{align}

\noindent The latter expression is an approximation of the Riemann integral

\begin{align}
&\int_0^s e^{-\frac{1}{\alpha}(t-t')}e^{-\frac{1}{\theta}(s-t')}\, dt' \nonumber\\
&\qquad = \frac{\alpha\theta}{\alpha+\theta} \left( e^{-\frac{1}{\alpha}(t-s)} - e^{-\frac{1}{\alpha}t}e^{-\frac{1}{\theta}s} \right),
\end{align}

\noindent which, in the stationary case $t\to\infty$ and $s\to\infty$, yields $\frac{\alpha\theta}{\alpha+\theta} e^{-\frac{1}{\alpha}(t-s)}$. All in all,

\begin{align}
\langle I_{0,t}^{t,\alpha}I_{0,s}^{s,\theta}\rangle = \frac{\alpha\theta}{\alpha+\theta} e^{-\frac{1}{\alpha}(t-s)}.
\end{align}

Consequently, for the covariance function we obtain

\begin{align}
\text{Cov}(X_t, X_s) = \frac{\alpha^2\beta\theta}{2(\alpha+\theta)(\alpha-\theta)} \left\lbrack \alpha e^{-\frac{1}{\alpha}(t-s)} - \theta e^{-\frac{1}{\theta}(t-s)} \right\rbrack.
\end{align}

From this, we can infer that

\begin{align}
\text{Var}(X_t) = \text{Cov}(X_t, X_t) = \frac{\alpha^2\beta\theta}{2(\alpha+\theta)}.
\end{align}

Further, the autocorrelation function $\text{ACF}(\tau)$, which is the normed autocovariance function ($\text{ACF}(\tau) = \text{Cov}(X_{\tau},X_0)/\text{Var}(X_0)$), reads

\begin{align}
\text{ACF}(\tau) = \frac{1}{\alpha-\theta} \left\lbrack \alpha e^{-\frac{1}{\alpha}\tau} - \theta e^{-\frac{1}{\theta}\tau} \right\rbrack.
\end{align}

In the case $\alpha=\theta$, we have  (again we write $X_t$ instead of $X_t^{(1)}$ and $W_t$ instead of $W_t^{(1)}$)

\begin{align}
\text{Cov}(X_t,X_s) = \frac{\beta}{\alpha} \bigl\langle J_{0,t}^t J_{0,s}^s \bigr\rangle,
\end{align}

\noindent where

\begin{align}
J_{a,b}^c := \int_a^b (c-t')e^{-\frac{1}{\alpha}(c-t')}\, dW_{t'}.
\end{align}

With similar calculations as in the latter case, we obtain

\begin{align}
\text{Cov}(X_t,X_s) &= \frac{\beta}{\alpha} \bigl\langle J_{0,s}^t J_{0,s}^s \bigr\rangle \nonumber\\
&= \frac{\beta}{\alpha} \int_0^s (t-t')e^{-\frac{1}{\alpha}(t-t')}(s-t')e^{-\frac{1}{\alpha}(s-t')}\, dt'\nonumber\\
&= \frac{\alpha^2\beta}{4} e^{-\frac{1}{\alpha}(t-s)} + \frac{\alpha\beta}{4} e^{-\frac{1}{\alpha}(t-s)}(t-s)
\end{align}

\noindent for the stationary case. From this, we can infer the relations

\begin{align}
\text{Var}(X_t) &= \frac{\alpha^2\beta}{4}\\
\text{ACF}(\tau) &= e^{-\frac{1}{\alpha}\tau} + \frac{1}{\alpha}\tau e^{-\frac{1}{\alpha}\tau}.
\end{align}

\section{Taylor length $\Lambda$ and correlation length $L$ of the linear model\label{app:B}}

As introduced in Sec.~\ref{sec:Optimization}, we obtain the Taylor length by a second order series expansion of the autocorrelation function $\text{ACF}(\tau)$ (cf. Appendix~\ref{app:AB}) at $\tau=0$ and the correlation length $L$ by its integral. First, we regard the case $\alpha\ne\theta$. From

\begin{align}
\text{ACF}(\tau) &= \frac{1}{\alpha-\theta} \left\lbrack \alpha e^{-\frac{1}{\alpha}\tau} - \theta e^{-\frac{1}{\theta}\tau} \right\rbrack \nonumber\\
&\approx \frac{1}{\alpha-\theta} \Bigl\lbrack \alpha(1-\frac{1}{\alpha}\tau+\frac{1}{2\alpha^2}\tau^2)\nonumber\\
&\qquad\qquad - \theta(1-\frac{1}{\theta}\tau+\frac{1}{2\theta^2}\tau^2) \Bigr\rbrack \nonumber\\
&= 1 - \frac12\frac{\tau^2}{\alpha\theta},
\end{align}

\noindent we obtain

\begin{align}
\Lambda = \sqrt{\alpha\theta}. \label{TaylorLength}
\end{align}

\noindent The correlation length $L$ can be calculated as follows:

\begin{align}\label{CorrLength}
L &= \int_0^{\infty} \text{ACF}(\tau)\, d\tau \nonumber\\ &= \frac{1}{\alpha-\theta} \left\lbrack \alpha\int_0^{\infty}e^{-\frac{1}{\alpha}\tau}\, d\tau - \theta\int_0^{\infty}e^{-\frac{1}{\theta}\tau}\, d\tau \right\rbrack \nonumber\\ &= \frac{\alpha^2 - \theta^2}{\alpha - \theta} \nonumber\\ &= \alpha + \theta
\end{align}

Similarly, we obtain $\Lambda = \alpha$ and $L=2\alpha$ in the case $\alpha=\theta$.

\end{document}